\documentclass{amsart}

\newtheorem*{theorem1}{Theorem 1}
\newtheorem*{theorem2}{Theorem 2}
\newtheorem*{theorem3}{Theorem 3}

\numberwithin{equation}{section}

\newcommand{\integers}{{\mathbb Z}}

\begin{document}

\title[Growth of Hyperbolic Gravitational Instantons]{On the Growth of the Number of 
Hyperbolic Gravitational Instantons with respect to Volume}

\author{John G. Ratcliffe}
\author{Steven T. Tschantz}

\address{Department of Mathematics, Vanderbilt University, Nashville, Tennessee 37240}


\date{}


\begin{abstract}
In this paper, we show that the number of hyperbolic gravitational 
instantons grows superexponentially with respect to volume. 
As an application, we show that the Hartle-Hawking 
wave function for the universe is infinitely peaked 
at a certain closed hyperbolic 3-manifold.  

\vspace{.1in}
\end{abstract}

\maketitle
\section{Introduction}

One of the standard models of the quantum origin of the universe 
begins with a {\it real tunneling geometry} 
consisting of a compact, connected, oriented, Riemannian 4-manifold $M_R$ 
and a Lorentzian 4-manifold $M_L$ joined across a totally geodesic 
spacelike hypersurface $\Sigma$ which serves as an initial Cauchy 
surface for the Lorentzian spacetime $M_L$. 
Given this setup one may pass to the double $2M_R=M_R^+ \cup M_R^-$ 
by joining two copies of $M_R$ across $\Sigma$.  
This is a closed, connected, oriented, Riemannian 4-manifold $M = 2M_R$ 
with a reflective isometry that fixes the totally geodesic submanifold $\Sigma$ 
and interchanges the two portions $M_R^\pm$. 
If $M$ is an Einstein manifold (constant Ricci curvature), 
then we call $M$ a {\it gravitational instanton}. 
If $M$ has constant negative sectional curvature equal to $-1$, 
then we call $M$ a {\it hyperbolic gravitational instanton}. 
As a reference for the theory of real tunneling geometries, 
see the paper by Gibbons and Hartle [6] 
and the papers by Gibbons [4], [5].

In our paper [9], we constructed an example of a hyperbolic gravitational 
instanton with a connected initial hypersurface $\Sigma$. 
In this paper, we show that the number of hyperbolic gravitational   
instantons, with initial hypersurface $\Sigma$,  
grows superexponentially with respect to volume. 
This implies that the Hartle-Hawking wave function for the universe is 
infinitely peaked at the hyperbolic 3-manifold $\Sigma$. 
In [1], Carlip showed that the 
$(2+1)$-dimensional Hartle-Hawking wave function is 
infinitely peaked at a certain hyperbolic 2-manifold.  
The results of our paper answer Carlip's question in [1] 
as to whether such results can be extended to $3+1$ dimensions. 

\section{Implications of superexponential growth} 

In the Hartle-Hawking approach to quantum gravity [7], 
the wave function of the universe, in the absence of matter, 
is specified by a sum-over-histories of the form 
$$\Psi[\Sigma,h] =\sum_M\int{\rm exp}(-I[M,g])dg$$
where $\Sigma$ is a closed orientable 3-manifold with Riemannian metric $h$, 
the sum is taken over all compact smooth 4-manifolds $M$ with $\partial M = \Sigma$, 
the integral is integrated over all Riemannian metrics $g$ on $M$ 
such that the Riemannian manifold $(\Sigma, h)$ is the boundary of $(M,g)$. 
Here $I[M,g]$ is the Euclidean action for the Riemannian manifold $(M,g)$,  
including a cosmological constant $\Lambda$, given by 
$$I[M, g] = -\frac{1}{16\pi}\int_M(R[g]-2\Lambda)\sqrt{g}\,d^4x - 
\frac{1}{8\pi}\int_\Sigma K[h]\sqrt{h}\,d^3x$$
where $R[g]$ is the scalar curvature of $g$ 
and $K[h]$ is the trace of the extrinsic curvature of $(\Sigma,h)$ in $(M,g)$. 
The sum $\Psi[\Sigma,h]$ gives the amplitude for $(\Sigma, h)$ 
to be the spatial geometry of the universe at the beginning of time. 

A local extremum of the action $I[M,g]$ is an Einstein 4-manifold $(M,g)$,  
with Ricci tensor $R_{\mu\nu} = \Lambda g_{\mu\nu}$,    
such that $\Sigma = \partial M$ is totally geodesic, see Carlip [2]. 
The double of a connected, orientable, Einstein 4-manifold, with 
a totally geodesic boundary, is a gravitational instanton. 
Therefore gravitational instantons correspond to local extrema 
of the Hartle-Hawking wave function.

The Hartle-Hawking wave function $\Psi$ is evaluated perturbatively 
by expanding around its local extrema in a saddle-point approximation 
$$\Psi[\Sigma,h] \approx \sum_{(M,g)} \Delta[M,g]{\rm exp}(-I[M,g])$$ 
where the sum is over all Einstein 4-manifolds $(M,g)$  
with totally geodesic boundary $\partial M=\Sigma$. 
The prefactor $\Delta[M,g]$ is a combination of determinants coming 
from gauge-fixing and small fluctuations around the extrema. 
The scalar curvature of an Einstein 4-manifold $(M,g)$,  
with Ricci tensor $R_{\mu\nu} = \Lambda g_{\mu\nu}$, 
is $R = 4\Lambda$, and so the action of $(M,g)$ is
$$I(M,g) = -\frac{\Lambda}{8\pi}{\rm Vol}(M,g).$$

We are interested in the case $\Lambda < 0$ and $(M,g)$ 
has constant negative sectional curvature $k = \Lambda/3$. 
In order to work with hyperbolic manifolds,  
we shall normalize curvature so that $k = -1$ by setting $\Lambda = -3$. 
In this paper, we shall prove the following theorem. 

\begin{theorem1} 
There is a closed, connected, orientable, hyperbolic 3-manifold $\Sigma$ 
such that the number of compact, connected, orientable, hyperbolic 4-manifolds $N$  
of volume at most $V$, with totally geodesic boundary $\partial N = \Sigma$, 
grows superexponentially with $V$. 
\end{theorem1}

Let $\Sigma$ and $N$ be as in the above theorem. 
Then $N$ is an Einstein 4-manifold 
with a totally geodesic boundary $\partial N = \Sigma$.  
Hence $N$ contributes to the saddle-point approximation of $\Psi[\Sigma, h]$. 
According to Carlip [3], the prefactor $\Delta[N,g]$ decays  
at most exponentially with ${\rm Vol}(N,g)$. 
Consequently Theorem 1 implies that the number of terms in 
in the saddle-point approximation for $\Psi[\Sigma,h]$ 
grows much faster than the exponential decay of the terms; 
therefore the sum in the saddle-point approximation 
$$\Psi[\Sigma,h] \approx \sum_{(M,g)}\Delta[M,g]{\rm exp}(-I[M,g])$$ 
diverges to $+\infty$. 
Thus the Hartle-Hawking wave function $\Psi$ 
is infinitely peaked at the hyperbolic 3-manifold $\Sigma$. 

\section{A hyperbolic gravitational instanton}

In our paper [9], we described a nonorientable closed hyperbolic 4-manifold $M_2$ 
that is obtained by gluing together the sides of a regular 120-cell 
with dihedral angle $2\pi/5$ in hyperbolic 4-space.  
The hyperbolic 4-manifold $M_2$ has a reflective symmetry along 
a totally geodesic nonorientable hypersurface $\Sigma_2$ 
that separates $M_2$ into two components. 
The orientable double cover of $M_2$ is a hyperbolic gravitational instanton 
$M$ with a reflective symmetry along a totally geodesic 
hypersurface $\Sigma$ that separates $M$ into two components. 
The hyperbolic 3-manifold $\Sigma$ is the orientable double cover of $\Sigma_2$. 
The Euler characteristic of $M$ is 52 and the volume of $\Sigma$ is approximately 
204.5. 

The Davis hyperbolic 4-manifold is a closed orientable 
hyperbolic 4-manifold that is obtained by gluing together 
the opposite sides of a regular 120-cell, 
with dihedral angle $2\pi/5$ in hyperbolic 4-space, 
by parallel translations. 
In our paper on the Davis hyperbolic 4-manifold [10], 
we described a regular 120-cell $P$, with center $(0,0,0,0,1)$,  
in the hyperboloid model of hyperbolic 4-space $H^4$ 
in Lorentzian 5-space $E^{(4,1)}$. 
In [10], we gave 120 Lorentzian $5\times 5$ matrices $A_i$, with $i=1,\ldots,120$ 
that represent the side-pairing maps of $P$ that glue up the Davis hyperbolic 
4-manifold. We now order the sides of $P$ so that the $i$th side of $P$ is 
$$S_i = (A_i^{-1}P)\cap P \ \ \hbox{for}\ \ i=1,\ldots,120.$$
Then 
$A_iS_i = S_{121-i} = -S_i$
and 
$$-S_i=P\cap A_iP \ \ \hbox{for}\ \ i=1,\ldots,120.$$
The sides of $P$ whose centers have coordinates $x_2=x_3=x_4=0$ 
have indices 1 and 120. 
The sides of $P$ whose centers have coordinate $x_1=0$
have indices $46,\ldots, 75$. 

In [9], we gave a geometric description of the side-pairing maps 
of a regular 120-cell that glue up the hyperbolic 4-manifold $M_2$. 
We now give a more explicit description of these side-pairing maps 
in terms of the side-pairing maps of the regular 120-cell $P$ in [10] 
that glue up the Davis hyperbolic 4-manifold.  
To be consistent with [10], we reverse the order of the coordinates in [9]  
and call the hyperplane $x_1=0$ the {\it equatorial plane}. 
Then side 1 is the {\it side at the north pole} 
and side 120 is the {\it side at the south pole}. 

We represent the side-pairing maps for $P$ that glue up the manifold $M_2$ 
by 120 Lorentzian $5\times 5$ matrices $B_i$  
such that $B_i=A_i$ for $i= 1, 46, 59, 60, 61, 62, 75, 120$, 
$B_i=D_1A_i$, where $D_1={\rm diag}(-1,1,1,1,1)$ represents 
the reflection in the $x_1=0$ hyperplane,  
for $i =$ 6, 7, 8, 9, 22, 23, 24, 25, 38, 39, 40, 41, 80, 81, 82, 
83, 96, 97, 98, 99, 112, 113, 114, 115, $B_i=D_2A_i$,   
where $D_2={\rm diag}(1,-1,1,1,1)$ represents 
the reflection in the $x_2=0$ hyperplane, 
for $i=$ 47, 48, 49, 50, 51, 52, 53, 54, 55, 56, 57, 58, 63, 64, 
65, 66, 67, 68, 69, 70, 71, 72, 73, 74, and $B_i = D_1D_2A_i$ for   
the remaining values of $i$. 

The side $S_i$ is paired to the side $S_j$ with $B_iS_i=S_j$. 
The pairs $i,j$ are listed in Table 1 with  $i$ over $j$. 
For example, the side $S_1$ at the north pole is paired to 
the side $S_{120}$ at the south pole. 
By Poincar\'e's fundamental polyhedron theorem, 
see Theorem 11.2.2 of [8], 
the fundamental group of $M_2$ has a presentation with 120 generators 
$y_i$ corresponding to the 120 side-pairing maps represented by $B_i$, 
120 side-pairing relations $y_iy_j=1$  
where the pairs $i,j$ are listed in Table 1 with $i$ over $j$, 
and 144 ridge cycle relations $y_iy_jy_ky_\ell y_m=1$ 
where $(i,j,k,\ell,m)$ is one of the 5-tuples listed in Table 2. 

As described in [9] the hypersurface $\Sigma_2$ of $M_2$ 
is obtained by gluing the sides of $P$ at the north and south poles 
to the equatorial cross-section of $P$. 
The hypersurface $\Sigma$ of $M$ is the orientable double cover 
of the nonorientable hyperbolic 3-manifold $\Sigma_2$. 
Let $\rho:\Sigma \to \Sigma$ be the covering transformation 
of the double covering of $\Sigma_2$ by $\Sigma$. 
Then $\rho$ is an orientation reversing isometry of $\Sigma$ of order 2. 

\begin{table} 
{
\small
\tabskip 0pt
\halign to\linewidth{\tabskip 0pt plus 1fil%
\hfil#&\hfil#&\hfil#&\hfil#&\hfil#%
&\hfil#&\hfil#&\hfil#&\hfil#&\hfil#%
&\hfil#&\hfil#&\hfil#&\hfil#&\hfil#%
&\hfil#&\hfil#&\hfil#&\hfil#&\hfil#%
\tabskip 0pt\cr
1&2&3&4&5&6&7&8&9&10&11&12&13&14&15&16&17&18&19&20\cr
120&3&2&5&4&9&8&7&6&11&10&13&12&15&14&19&18&17&16&21\cr
\noalign{\vskip 10pt}
21&22&23&24&25&26&27&28&29&30&31&32&33&34&35&36&37&38&39&40\cr
20&25&24&23&22&27&26&31&30&29&28&33&32&35&34&37&36&41&40&39\cr
\noalign{\vskip 10pt}
41&42&43&44&45&46&47&48&49&50&51&52&53&54&55&56&57&58&59&60\cr
38&43&42&45&44&75&50&49&48&47&54&53&52&51&58&57&56&55&62&61\cr
\noalign{\vskip 10pt}
61&62&63&64&65&66&67&68&69&70&71&72&73&74&75&76&77&78&79&80\cr
60&59&66&65&64&63&70&69&68&67&74&73&72&71&46&77&76&79&78&83\cr
\noalign{\vskip 10pt}
81&82&83&84&85&86&87&88&89&90&91&92&93&94&95&96&97&98&99&100\cr
82&81&80&85&84&87&86&89&88&93&92&91&90&95&94&99&98&97&96&101\cr
\noalign{\vskip 10pt}
101&102&103&104&105&106&107&108&109&110&111&112&113&114&115&116&117&118&119&120\cr
100&105&104&103&102&107&106&109&108&111&110&115&114&113&112&117&116&119&118&1\cr
}
}
\vspace{.2in}
\caption{Table of indices for the paired sides of the 120-cell P 
that glue up the hyperbolic 4-manifold $M_2$.}
\end{table}

\begin{table}
{
\small
\tabskip 0pt
\halign to\linewidth{
\tabskip 0pt(#&\tabskip 3pt\hfil#,&\hfil#,&\hfil#,&\hfil#,&\hfil#)
\tabskip 3pt plus 1 fil&
\tabskip 0pt(#&\tabskip 3pt\hfil#,&\hfil#,&\hfil#,&\hfil#,&\hfil#)
\tabskip 3pt plus 1 fil&
\tabskip 0pt(#&\tabskip 3pt\hfil#,&\hfil#,&\hfil#,&\hfil#,&\hfil#)
\tabskip 3pt plus 1 fil&
\tabskip 0pt(#&\tabskip 3pt\hfil#,&\hfil#,&\hfil#,&\hfil#,&\hfil#)
\tabskip 0pt\cr
&1&2&35&76&109&&7&30&72&32&13&&46&47&58&63&74&&59&69&72&71&70\cr
&1&3&34&77&108&&10&12&31&63&25&&46&48&57&64&73&&59&70&71&72&69\cr
&1&4&37&78&111&&10&13&29&65&23&&46&49&56&65&72&&59&82&97&96&83\cr
&1&5&36&79&110&&10&23&65&29&13&&46&50&55&66&71&&59&83&96&97&82\cr
&1&6&41&80&115&&10&25&63&31&12&&46&76&95&100&87&&59&95&108&108&95\cr
&1&7&40&81&114&&12&12&27&59&27&&46&77&94&101&86&&59&101&118&118&101\cr
&1&8&39&82&113&&14&14&37&60&37&&46&88&111&116&107&&60&64&69&67&66\cr
&1&9&38&83&112&&14&17&40&52&35&&46&89&110&117&106&&60&66&67&69&64\cr
&1&10&43&84&117&&14&19&38&54&34&&47&48&53&59&54&&60&79&88&88&79\cr
&1&11&42&85&116&&14&34&54&38&19&&47&54&59&53&48&&60&85&106&106&85\cr
&1&12&45&86&119&&14&35&52&40&17&&47&77&91&98&79&&60&97&114&112&99\cr
&1&13&44&87&118&&16&21&39&57&37&&47&79&98&91&77&&60&99&112&114&97\cr
&2&2&21&59&21&&16&25&36&48&35&&47&89&109&112&93&&63&83&100&104&85\cr
&2&4&25&55&19&&16&35&48&36&25&&47&93&112&109&89&&63&85&104&100&83\cr
&2&5&24&56&18&&16&37&57&39&21&&48&77&90&99&78&&63&99&116&118&105\cr
&2&8&17&49&15&&17&21&38&58&36&&48&78&99&90&77&&63&105&118&116&99\cr
&2&9&16&50&14&&17&24&37&47&35&&48&88&109&113&92&&64&82&100&105&84\cr
&2&14&50&16&9&&17&35&47&37&24&&48&92&113&109&88&&64&84&105&100&82\cr
&2&15&49&17&8&&17&36&58&38&21&&51&53&56&60&58&&64&98&117&118&104\cr
&2&18&56&24&5&&20&27&44&46&35&&51&58&60&56&53&&64&104&118&117&98\cr
&2&19&55&25&4&&20&35&75&44&27&&51&77&89&90&83&&67&83&102&107&87\cr
&4&7&21&51&19&&22&23&40&59&41&&51&83&90&89&77&&67&87&107&102&83\cr
&4&9&20&53&17&&22&31&44&73&43&&51&93&110&113&95&&67&101&113&116&105\cr
&4&11&32&46&15&&22&41&59&40&23&&51&95&113&110&93&&67&105&116&113&101\cr
&4&15&75&32&11&&22&43&73&44&31&&52&77&88&91&82&&68&82&103&106&87\cr
&4&17&53&20&9&&23&30&44&74&42&&52&82&91&88&77&&68&87&106&103&82\cr
&4&19&51&21&7&&23&42&74&44&30&&52&92&111&112&95&&68&101&112&117&104\cr
&6&8&23&60&25&&26&30&43&63&41&&52&95&112&111&92&&68&104&117&112&101\cr
&6&11&30&68&27&&26&31&42&64&40&&55&79&92&94&83&&71&85&98&103&87\cr
&6&13&33&71&31&&26&40&64&42&31&&55&83&94&92&79&&71&87&103&98&85\cr
&6&25&60&23&8&&26&41&63&43&30&&55&93&108&110&99&&71&105&112&119&107\cr
&6&27&68&30&11&&28&33&45&67&41&&55&99&110&108&93&&71&107&119&112&105\cr
&6&31&71&33&13&&28&41&67&45&33&&56&78&93&94&82&&72&84&99&102&87\cr
&7&10&31&67&27&&29&32&45&68&40&&56&82&94&93&78&&72&87&102&99&84\cr
&7&13&32&72&30&&29&40&68&45&32&&56&92&108&111&98&&72&104&113&119&106\cr
&7&27&67&31&10&&32&32&43&60&43&&56&98&111&108&92&&72&106&119&113&104\cr}
}
\vspace{.2in}
\caption{Table of indices for the ridge cycle relators 
for the fundamental group of the hyperbolic 4-manifold $M_2$.}
\end{table}

\section{A 4-dimensional pair of pants}

In this section, we construct a compact, connected, orientable, 
hyperbolic 4-manifold $Y$ 
with a totally geodesic boundary equal to the disjoint union 
of three copies of the hyperbolic 3-manifold $\Sigma$ defined 
in the previous section. 

Consider the homomorphism $\eta: F \to \integers/3\integers$ 
from the free group $F$ with generators 
$y_1,\ldots,y_{120}$ onto the group $\integers/3\integers$ of integers module 3  
so that the generators that map to 1 mod 3 have indices
3, 6, 7, 13, 18, 19, 21, 27, 30, 31, 35, 45, 77, 87, 92,
   93, 95, 101, 104, 105, 109, 112, 113, 119, 
the generators that map to $-1$ mod 3 have indices
2, 8, 9, 12, 16, 17, 20, 26, 28, 29, 34, 44, 76, 86, 90,
   91, 94, 100, 102, 103, 108, 114, 115, 118, 
and all other generators map to 0 mod 3. 
One can check that all 120 side-pairing relators $y_iy_j$ 
and all 144 ridge cycle relators $y_iy_jy_ky_\ell y_m$ 
for the presentation of the fundamental group of $M_2$ 
are in the kernel of $\eta$. 
Hence $\eta$ induces a homomorphism $\overline\eta: G \to \integers/3\integers$  
from the fundamental group $G$ of $M_2$ onto $\integers/3\integers$. 
This implies that $M_2$ has a regular triple cover $M_3$ 
corresponding to the kernel of $\overline\eta$. 

Let $p:M_3 \to M_2$ be the regular covering projection. 
The hyperbolic 4-manifold $M_3$ is nonorientable, 
since $M_2$ is nonorientable and the degree of $p$ is odd. 
Let $\Sigma_3 = p^{-1}(\Sigma_2)$. 
Then $\Sigma_3$ is a totally geodesic hypersurface of $M_3$ 
that separates $M_3$ into two components, 
since $\Sigma_2$ is a totally geodesic hypersurface of $M_2$ 
that separates $M_2$ into two components. 
The hyperbolic 4-manifold $M_3$ has a reflective symmetry along $\Sigma_3$, 
since $M_2$ has a reflective symmetry along $\Sigma_2$   
and the covering projection $p:M_3 \to M_2$ is regular.  

The homomorphism $\overline\eta: G \to \integers/3\integers$ 
maps the generators $g_1$ and $g_{120}$, 
corresponding to the sides of $P$ at the north and south poles, 
and the generators $g_{46},\ldots,g_{75}$, corresponding 
to the sides of $P$ centered on the equatorial plane, to $0$ mod 3. 
This implies the hypersurface $\Sigma_2$ is evenly covered by $p:M_3 \to M_2$. 
Therefore the hyperbolic 3-manifold $\Sigma_3$ 
is the disjoint union of three copies of $\Sigma_2$ 
with $p: M_3 \to M_2$ mapping each copy isometrically onto $\Sigma_2$.  

Let $H$ be the subgroup of $G$ corresponding to the cover $M_3$. 
Then $H$ is a normal subgroup of $G$ of index three, 
since $H = {\rm ker}(\overline\eta)$. 
The orientable double cover $M$ of $M_2$ corresponds to the subgroup $G_0$ of $G$ 
of orientation preserving elements. 
The orientable double cover $\tilde M$ of $M_3$ corresponds 
to the subgroup $H_0$ of $H$ of orientation preserving elements. 
By the Diamond Isomorphism Theorem $H_0 = H\cap G_0$ is a normal subgroup of $G_0$ 
and 
$$G_0/H_0 = G_0/H\cap G_0 \cong  G_0H/H = G/H.$$
Therefore $\tilde M$ is a regular triple cover of $M$. 

Let $q:\tilde M \to M$ be the regular covering projection 
and let $\tilde\Sigma = q^{-1}(\Sigma)$. 
Then $\tilde\Sigma$ is a totally geodesic hypersurface of $\tilde M$ 
that separates $\tilde M$ into two components, 
since $\Sigma$ is a totally geodesic hypersurface of $M$ 
that separates $M$ into two components. 
The hyperbolic 4-manifold $\tilde M$ has a reflective symmetry along $\tilde\Sigma$, 
since $M$ has a reflective symmetry along $\Sigma$   
and the covering projection $q:\tilde M \to M$ is regular. 

Let $p_2:M\to M_2$ and $p_3:\tilde M \to M_3$ be 
the orientable double covering projections. 
Then $pp_3=p_2q$, since the composition of the inclusions 
$H_0 \subset H \subset G$ is the same as the composition of the inclusions 
$H_0 \subset G_0 \subset G$. 
Now observe that 
$$\tilde\Sigma = q^{-1}(\Sigma) = q^{-1}p_2^{-1}(\Sigma_2)=
	p_3^{-1}p^{-1}(\Sigma_2) = p_3^{-1}(\Sigma_3).$$
Hence $\tilde\Sigma$ double covers $\Sigma_3$. 
Now $\tilde\Sigma$ is an orientable hyperbolic 3-manifold, 
since $\Sigma$ is an orientable hyperbolic 3-manifold. 
Hence $\tilde\Sigma$ is the orientable double cover of $\Sigma_3$.   
Therefore the hyperbolic 3-manifold $\tilde\Sigma$ 
is the disjoint union of three copies of $\Sigma$ 
with $q: \tilde M \to M$ mapping each copy isometrically onto $\Sigma$.

Let $Y$ be the closure in $\tilde M$ of one of the connected components 
of the complement of $\tilde\Sigma$ in $\tilde M$. 
Then $Y$ is a compact, connected, orientable,  
hyperbolic 4-manifold with a totally geodesic boundary $\partial Y = \tilde\Sigma$. 
In particular $\partial Y$ has three components each isometric to $\Sigma$ 
via $q: \tilde M \to M$. 
In other words, $Y$ is a 4-dimensional hyperbolic pair of pants 
with ``waist" and ``cuffs" isometric to $\Sigma$.

\section{Superexponential growth of hyperbolic gravitational instantons} 

We are now ready to prove the main results of this paper. 
In the following theorem, we shall not distinguish between isometric manifolds 
when counting manifolds. 

\begin{theorem1} 
There is a closed, connected, orientable, hyperbolic 3-manifold $\Sigma$ 
such that the number of compact, connected, orientable, hyperbolic 4-manifolds $N$  
of volume at most $V$, with totally geodesic boundary $\partial N = \Sigma$,  grows 
superexponentially with $V$. 
\end{theorem1}

\begin{proof}
Let $Y$ be the 4-dimensional hyperbolic pair of pants 
described in the previous section.
Turn the pair of pants up-side-down and think of $Y$ as being in the shape of a ``Y". 
Each endpoint of ``Y" represents a component of $\partial Y$ isometric 
to the hyperbolic 3-manifold $\Sigma$ described in the previous section. 
We call the three components of $\partial Y$, 
the {\it left component} $A$, the {\it bottom component} $B$, 
and the {\it right component} $C$. 
The covering projection $q:\tilde M \to M$ restricts 
to isometries $\alpha:A\to\Sigma$, $\beta:B\to\Sigma$, $\gamma:C\to\Sigma$. 
We choose an orientation for $Y$ and orient $A, B, C$  
with the induced orientation from $Y$. 
The covering projection $q:\tilde M \to M$ has a covering transformation 
$\delta:\tilde M \to \tilde M$ that cyclically permutes $A, B, C$, 
since $q$ is regular. 
The isometry $\delta$ is orientation preserving, since $\delta$ has order three. 
Hence $\delta$ preserves orientation when cyclically permuting $A, B, C$. 
This implies that we can orient $\Sigma$ so that $\alpha,\beta,\gamma$ 
are all orientation preserving.

Let $n$ be a positive integer greater than one. 
Glue $n$-copies, $Y_1, \ldots Y_n$, of $Y$ together, 
so that the right component $C_i$ of $\partial Y_i$ is glued to 
the bottom component $B_{i+1}$ of $\partial Y_{i+1}$
by the orientation reversing isometry corresponding to $\rho:\Sigma \to \Sigma$ 
for each $i=1,\ldots, {n-1}$.  
More specifically, we glue $C_i$ to $B_{i+1}$ by identifying the point 
$y$ of $C_i$ with the point $\beta_{i+1}^{-1}\rho\gamma_i(y)$ of $B_{i+1}$,  
where $\gamma_i:C_i\to\Sigma$ and $\beta_{i+1}:B_{i+1}\to\Sigma$ 
are the isometries that correspond to $\gamma$ and $\beta$, respectively. 
This gives a compact, connected, oriented, hyperbolic 4-manifold $N_n$, 
with a totally geodesic boundary, 
such that $\partial N_n$ is the disjoint union of $n+2$ copies of $\Sigma$, 
labeled $\Sigma_0,\Sigma_1, \ldots, \Sigma_{n+1}$ 
where $\Sigma_0$ corresponds to the bottom component of $\partial Y_1$, 
and $\Sigma_i$ corresponds to the left component of $\partial Y_i$ for 
$i=1,\ldots, n$, and $\Sigma_{n+1}$ corresponds to the right component 
of $\partial Y_n$. Moreover the natural injection of $Y_i$ into $N_n$ 
is orientation preserving for each $i$, 
since the gluing map $\beta_{i+1}^{-1}\rho\gamma_i:C_i \to B_{i+1}$ is 
orientation reversing for each $i$. 

Assume $n$ is odd and let $m=(n+1)/2$.  
Let $\sigma$ be a permutation of the set $\{1,2,\ldots,m\}$. 
We construct a compact, connected, oriented, hyperbolic 4-manifold $N_\sigma$   
from $N_n$ by gluing the boundary component $\Sigma_i$ 
to the boundary component $\Sigma_{m+\sigma(i)}$  
by the orientation reversing isometry corresponding to 
$\rho:\Sigma\to \Sigma$, for each $i=1,2,\ldots,m$. 
Then $N_\sigma$ has a totally geodesic boundary corresponding to 
the boundary component $\Sigma_0$ of $N_n$. 
To emphasize that the boundary of $N_\sigma$ is isometric to $\Sigma$, 
we shall identify $\partial N_\sigma$ with $\Sigma$. 

Let $\sigma$ and $\tau$ be permutations of the set $\{1,2,\ldots,m\}$ 
and let $\phi:N_\sigma \to N_\tau$ be an isometry 
that restricts to the identity map on $\partial N_\sigma = \Sigma = \partial N_\tau$. 
We now show that $\sigma=\tau$ and $\phi$ is the identity map. 
The manifolds $N_\sigma$ and $N_\tau$ are obtained from the manifold $N_n$ 
by gluing the boundary components $\Sigma_1, \ldots, \Sigma_{m}$ 
to the boundary components $\Sigma_{m+1},\ldots,\Sigma_{n+1}$ 
in the order specified by the permutations $\sigma$ and $\tau$. 
Let $N_n^\circ$ be the interior of $N_n$. 
Then $N_n^\circ$ is an open dense submanifold of both $N_\sigma$ and $N_\tau$. 
If an isometry $\psi$ of hyperbolic 4-space $H^4$ restricts 
to the identity map on a hyperplane of $H^4$, 
then either $\psi$ is the identity map or the reflection in the hyperplane. 
This implies that $\phi$ restricts to the identity map 
on an open regular neighborhood of 
$\partial N_\sigma = \Sigma$ in $\Sigma\cup N_n^\circ$. 
Therefore $\phi$ restricts to the identity map on all 
of $\Sigma\cup N_n^\circ$, since $\phi$ is analytic. 
This implies that $\sigma=\tau$ and $\phi$ is the identity map, 
since $\phi$ is continuous. 

Let $\tau_1$ and $\tau_2$ be permutations of the set $\{1,2,\ldots,m\}$   
and let $\phi_i:N_\sigma \to N_{\tau_i}$, for $i=1,2$, be isometries 
such that $\phi_1|\partial N_\sigma = \phi_2|\partial N_\sigma$.  
Then $\phi_2\phi_1^{-1}:N_{\tau_1} \to N_{\tau_2}$ is an isometry 
which restricts to the identity map on $\partial N_\sigma = \Sigma$. 
By the previous argument, $\tau_1=\tau_2$ and $\phi_2\phi_1^{-1}$ 
is the identity map. Therefore $\phi_1=\phi_2$. 
Thus an isometry $\phi:N_\sigma\to N_\tau$ depends only 
on the permutation $\sigma$ and the isometry $\phi|\partial N_\sigma$ 
of $\partial N_\sigma = \Sigma$.

The manifold $\Sigma$ has a finite number $k$ of isometries, 
since $\Sigma$ is a closed hyperbolic 3-manifold. 
It now follows that for each permutation $\sigma$ 
of the set $\{1,2,\ldots,m\}$, there are at most $k$ 
permutations $\tau$ of $\{1,2,\ldots,m\}$ such that 
$N_\sigma$ is isometric to $N_\tau$.
Hence there are at least $m!/k$ nonisometric 
hyperbolic manifolds $N_\sigma$.

Now by Stirling's Formula
$$m! > \sqrt{2\pi m}\left(\frac{m}{e}\right)^m.$$
Therefore  $m!/k$ grows superexponentially with $n$, since $m=(n+1)/2$.  
Now
$${\rm Vol}(N_\sigma) = {\rm Vol}(N_n) = n{\rm Vol}(Y).$$
Thus the number of compact, connected, orientable, hyperbolic 4-manifolds $N$ 
of volume at most $V$, with totally geodesic boundary $\partial N = \Sigma$, 
grows superexponentially with $V$. 

\end{proof}

A {\it marked gravitation instanton} is a pair $(M,\Sigma)$ 
consisting of a gravitational instanton $M$ and the fixed space $\Sigma$ 
of its reflection symmetry $\theta$. 
The choice of an initial hypersurface $\Sigma$ (or 
equivalently a reflection symmetry $\theta$) of a gravitational 
instanton $M$ is called a {\it marking} of $M$. 
A gravitational instanton $M$ has only finitely many markings,  
since the group of isometries of a closed hyperbolic manifold is finite.  
Two marked gravitational instantons $(M,\Sigma)$ and $(M',\Sigma')$ 
are said to be {\it isometric} if there is an isometry 
$\phi:(M,\Sigma) \to (M',\Sigma')$ of pairs of manifolds, 
that is, an isometry $\phi:M\to M'$ such that $\phi(\Sigma)=\Sigma'$.  
We shall not distinguish between isometric manifolds (resp. isometric 
pairs of manifolds) when counting gravitational instantons  
(resp. marked gravitational instantons). 
Theorem 1 is equivalent to the following theorem. 

\begin{theorem2} 
There is a closed, connected, orientable, hyperbolic 3-manifold $\Sigma$ 
such that the number of marked hyperbolic gravitational instantons $(M,\Sigma)$   
of volume at most $V$ grows superexponentially with $V$.
\end{theorem2}
 
The next theorem improves Theorem 2 in a subtle but nontrivial way.  

\begin{theorem3} 
There is a closed, connected, orientable, hyperbolic 3-manifold $\Sigma$ 
such that the number of hyperbolic gravitational instantons $M$ 
of volume at most $V$, with an initial hypersurface isometric to $\Sigma$,  
grows superexponentially with $V$. 
\end{theorem3}

\begin{proof}
Let $\Sigma$ be as in Theorem 1 and let $Y$ be as in the proof of Theorem 1. 
Let $S$ be the compact hyperbolic 4-manifold obtained by doubling $Y$ 
along the union of the right and left components of $\partial Y$. 
Then $S$ has a totally geodesic boundary 
consisting of two copies of $\Sigma$ corresponding to two copies of 
the bottom component of $\partial Y$. 
We call $S$ a 4-dimensional hyperbolic sleeve.  
By attaching enough sleeves end-to-end to the boundary components of $Y$, 
we may assume without loss of generality 
that the diameter of $\Sigma$ is less than the distance between 
the components of $\partial Y$. 

By the proof of Theorem 1, 
the number of compact, connected, orientable, hyperbolic 
4-manifolds $N$ of volume at most $V$, 
with totally geodesic boundary $\partial N =\Sigma$, 
containing $Y$ as a submanifold such that the bottom component of 
$\partial Y$ is equal to $\partial N$, grows superexponentially with $V$. 

Now consider a hyperbolic gravitational instanton $M$ obtained 
by doubling one of the above 4-manifolds $N$. 
Then $M$ has $\Sigma$ as an initial hypersurface. 
Suppose $M$ has another initial hypersurface $\Sigma'$ isometric to $\Sigma$. 
Now $\Sigma$ subdivides $M$ into two isometric 
compact, connected, hyperbolic 4-manifolds $N$ and $N^{-}$ with boundary $\Sigma$. 
The submanifold $\Sigma'$ subdivides $M$ in a similar fashion. 
Neither $N$ nor $N^{-}$ is contained in the complement of $\Sigma'$ in $M$, 
since the volume of $N$ and $N^{-}$ is half the volume of $M$ 
and any compact connected subset of the complement of $\Sigma'$ in $M$ 
has volume less than half the volume of $M$. 
Therefore $\Sigma$ and $\Sigma'$ must intersect. 

Now since the diameter of $\Sigma'$ is less than the distance 
between the boundary components of $Y$, 
we conclude that $\Sigma'$ must be contained in the 
interior of the compact submanifold $X = Y \cup Y^{-}$ 
of $M$ where $Y^{-}$ is the reflection of $Y$ 
across its bottom boundary component $\Sigma$. 

It follows from Theorems 9.6.2, 12.1.7, and 12.1.15 of [8] 
that $\Sigma'$ contains three noncoplanar closed geodesics. 
Let $\ell$ be the length of the longest of these geodesics. 
Then $X$ contains only finitely many closed geodesics 
of length at most $\ell$, since $X$ is a compact hyperbolic manifold. 
Therefore there are only finitely many possibilities for $\Sigma'$, 
since $\Sigma'$ is spanned by any three noncoplanar closed geodesics in $\Sigma'$. 
Moreover, the number of possibilities for $\Sigma'$ 
is bounded by a number $m$ that depends only on $Y$. 
Therefore $M$ can be marked 
by an initial hypersurface isometric to $\Sigma$ 
in at most $m$ ways depending only on $Y$. 
Thus the proof of Theorem 1 implies that 
the number of hyperbolic gravitational instantons $M$ of volume at most $V$, 
with an initial hypersurface isometric to $\Sigma$,  
grows superexponentially with $V$. 
\end{proof}

In order to clarify the difference between Theorems 2 and 3, 
consider a marked hyperbolic gravitational instanton $(M,\Sigma)$ as in Theorem 2. 
The group of isometries of $M$ acts on the set of initial hypersurfaces of $M$ 
that are isometric to $\Sigma$. 
The number of orbits under this action is the number of marked 
gravitational instantons counted in Theorem 2 
whose underlying gravitational instanton is $M$, 
whereas $M$ is counted only once in Theorem 3.  
Thus the growth in Theorem 3 is a priori slower than the growth in Theorem 2 
but nevertheless the growth in Theorem 3 is still superexponential. 

\vspace{.2in}
\centerline{References}

\begin{enumerate}
\item Carlip, S., Entropy versus action in the (2+1)-dimensional 
Hartle-Hawking wave function, 
{\it Physical Rev. D}, 46 (1992), 4387-4395.

\item Carlip, S., Real tunnelling solutions and the Hartle-Hawking wavefunction, 
{\it Class. Quantum Grav.}, 10 (1993), 1057-1064.

\item Carlip, S., Dominant topologies in Euclidean quantum gravity, 
{\it Class. Quantum Grav.}, 15 (1998), 2629-2638.

\item Gibbons, G. W., Tunnelling with a negative cosmological constant, 
{\it Nucl. Phys. B}, 472 (1996), 683-708.

\item Gibbons, G. W., Real tunnelling geometries, 
{\it Class. Quantum Grav.}, 15 (1998) 2605-2612.

\item Gibbons, G. W. and Hartle, J. B., Real tunneling geometries 
and the large-scale topology of the universe, 
{\it Physical Rev. D}, 42 (1990), 2458-2468.

\item Hartle, J. B. and Hawking, S. W., Wave function of the universe,
{\it Physical Rev. D}, 28 (1983), 2960-2975.

\item Ratcliffe, J. G., {\it Foundations of Hyperbolic Manifolds}, 
Springer-Verlag, New York, 1994. 

\item Ratcliffe, J. G. and Tschantz, S. T., 
Gravitational instantons of constant curvature, 
{\it Class. Quantum Grav.}, 15 (1998) 2613-2627.

\item Ratcliffe, J. G. and Tschantz, S. T., 
On the Davis hyperbolic 4-manifold, 
{\it Topology Appl.} (2000), to appear.

\end{enumerate}
\end{document}